\documentclass[12pt,colorlinks=true,linkcolor=blue,citecolor=blue,urlcolor=blue]{article}
\makeatletter
\let\frontmatter@title@above=\relax
\makeatother
\usepackage{times}
\usepackage{geometry}
\usepackage[utf8]{inputenc}
\usepackage{enumitem,amssymb}
\usepackage{ragged2e}
\usepackage{setspace}
\usepackage{graphicx}
\usepackage{floatrow}
\usepackage{orcidlink}
\usepackage{makecell}
\usepackage{color}
\usepackage{style_aastex}
\usepackage[para]{threeparttablex}
\usepackage[bf,sf,small,compact]{titlesec}
\titlespacing*{\section}{0pt}{10pt}{6pt}
\titlespacing*{\subsection}{0pt}{3pt}{3pt}
\titlespacing*{\subsubsection}{0pt}{3pt}{3pt}
\newlist{thematic}{itemize}{8}
\setlist[thematic]{label=$\square$}
\usepackage{pifont}
\begin{document}
\pagenumbering{gobble}
\RaggedRight
%\noindent {\fontsize{16}{20} \selectfont White Paper for the 2024 Solar \& Space Physics Decadal Survey}
\begin{center}
%\title{Radio Studies of the Middle Corona}
{\fontsize{22}{35}\selectfont Radio Imaging Spectropolarimetry of \\ 
\vspace{0.2cm}
CMEs and CME Progenitors}

%\textit{\fontsize{16}{20}\selectfont Current State and New Prospects in the Next Decade}
\end{center}

%\vspace{0.3cm}

\normalsize

\justifying

%\begin{figure}[!ht]
%\floatbox[{\capbeside\thisfloatsetup{capbesideposition={right,top},capbesidewidth=4cm}}]{figure}[\FBwidth]
%{\includegraphics[width=1.0\textwidth]{title_image.pdf}}
%\end{figure}

%\noindent \textbf{Category:} Basic research\\
%\noindent \textbf{Topics:} Solar physics, Space weather needs

%\noindent \textbf{Secondary Category:} Space weather R2O2R loop \\
%\bigskip

% Principle Author
\noindent \textbf{Principal Author:} \\
Bin Chen$^{1}$ \orcidlink{0000-0002-0660-3350} \textit{New Jersey Institute of Technology} \\
Email: \href{mailto:binchen@njit.edu}{binchen@njit.edu}; Phone: (973) 596-3565; Web: \href{https://binchensun.org}{https://binchensun.org}

\smallskip

% Add Co-authors below
\noindent \textbf{Co-authors}\\
Timothy~S.~Bastian$^{2}$%
\orcidlink{0000-0002-0713-0604},
Sarah~Gibson$^{3}$%
\orcidlink{0000-0001-9831-2640},
Yuhong~Fan$^{3}$%
\orcidlink{0000-0003-1027-0795},
Stephen~M.~White$^{4}$%
\orcidlink{0000-0002-8574-8629},
Dale~E.~Gary$^{1}$%
\orcidlink{0000-0003-2520-8396},
Angelos Vourlidas$^5$%
\orcidlink{0000-0002-8164-5948},
Sijie~Yu$^{1}$%
\orcidlink{0000-0003-2872-2614}, 
Surajit~Mondal$^{1}$%
\orcidlink{0000-0002-2325-5298},
Gregory~D.~Fleishman$^{1}$%
\orcidlink{0000-0001-5557-2100}, 
Pascal Saint-Hilaire$^{6}$%
\orcidlink{0000-0002-8283-4556}

%Add Affiliations below
{\fontsize{11}{13}\selectfont \noindent 
[1] New Jersey Institute of Technology; 
[2] National Radio Astronomy Observatory;
[3] High Altitude Observatory;
[4] Air Force Research Laboratory; 
[5] JHU Applied Physics Laboratory;
[6] University of California, Berkeley
}

\noindent \textbf{Co-Signers \& Affiliations}: see spreadsheet

%\maketitle

\begin{figure}[!ht]
%\floatbox[{\capbeside\thisfloatsetup{capbesideposition={right,top},capbesidewidth=4.3cm}}]{figure}[\FBwidth]
\includegraphics[width=0.85\textwidth]{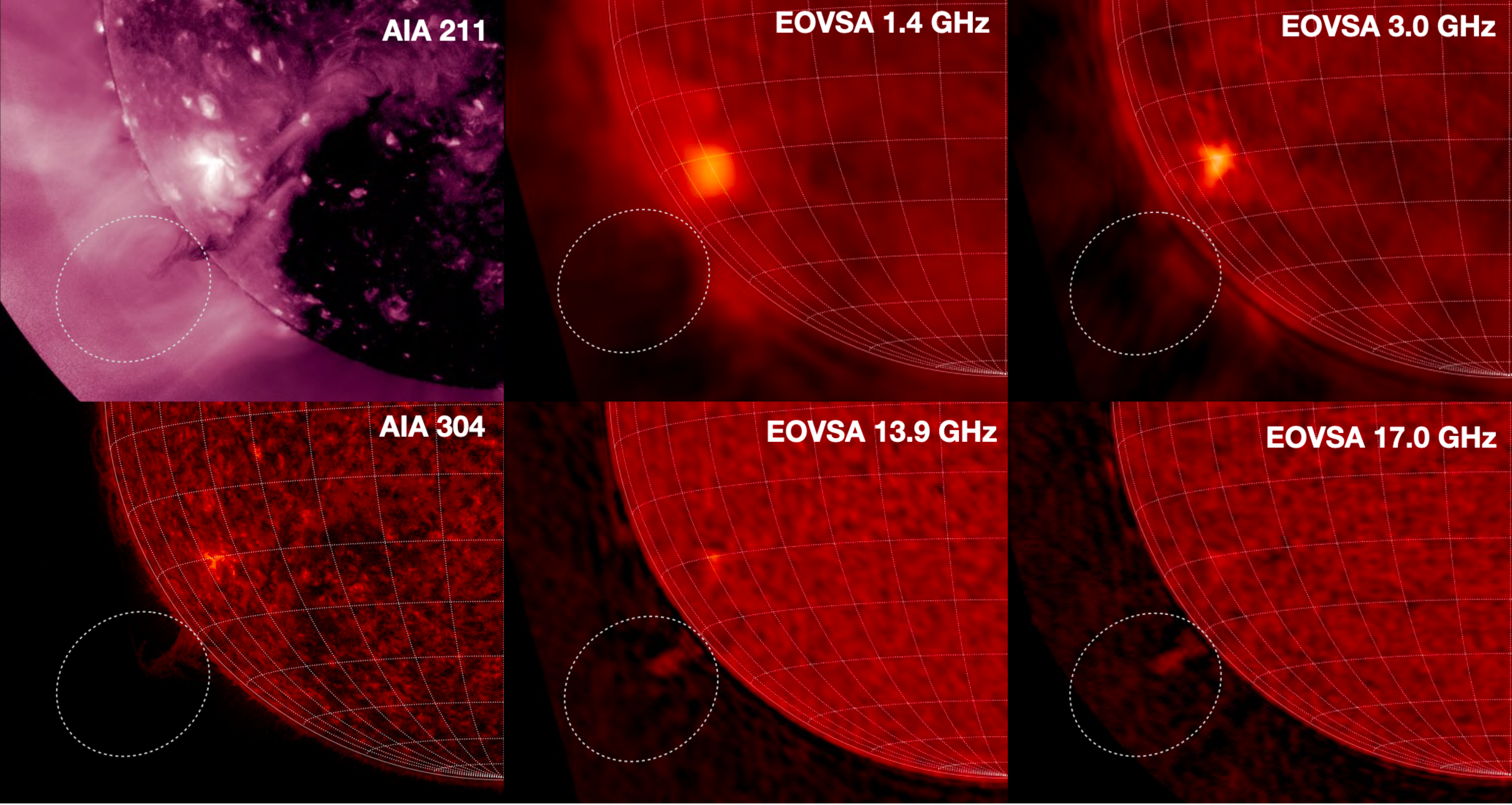}
%\caption{\fontsize{11}{13}\selectfont Multi-frequency radio images of a filament-hosting coronal cavity observed by the Expanded Owens Valley Solar Array. The images were obtained with a full day integration taking advantage of the Earth-rotational synthesis to improve the u-v coverage.}
\label{fig:cavity_eovsa}
\end{figure}

\vspace{-0.2cm}

\noindent \textbf{Synopsis} \\
Coronal mass ejections (CMEs) are the most important drivers of space weather. Central to most CMEs is thought to be the eruption of a bundle of highly twisted magnetic field lines known as magnetic flux ropes. A comprehensive understanding of CMEs and their impacts hence requires detailed observations of physical parameters that lead to the formation, destabilization, and eventual eruption of the magnetic flux ropes.  
Recent advances in remote-sensing observations of coronal cavities, filament channels, sigmoids, EUV ``hot channels,'' white light CMEs, and in situ observations of magnetic clouds points to the possibility of significant progress in understanding CMEs. In this white paper, we provide a brief overview of the potential of radio diagnostics for CMEs and CME progenitors, with a particular focus on the unique means for constraining their magnetic field and energetic electron population. Using synthetic observations based on realistic 3D MHD models, we also demonstrate the transformative potential of advancing such diagnostics by using broadband radio imaging spectropolarimetry with a high image dynamic range and high image fidelity. To achieve this goal, a solar-dedicated radio facility with such capabilities is recommended for implementation in the coming decade.

\newpage

\vspace{-2cm}

\pagenumbering{arabic}
\setcounter{page}{1}

\section{Introduction}\label{sec:intro}

Coronal mass ejections (CMEs) are violent eruptions of coronal plasma and magnetic field from the Sun, involving a few times $10^{14}$ to several $\times 10^{16}$ grams of the material moving into the interplanetary medium  with  speeds  of  a  few  hundred  to  more  than  2,000~km~s$^{-1}$. CMEs can drive significant geomagnetic storms and other space weather activity. In fact, the strongest space weather effects are usually related to the impact of CMEs with $B_z$ anti-aligned with the Earth's magnetic field. Therefore, a detailed understanding of the formation, initiation, eruption, and evolution of CMEs and their subsequent geospace impacts is of vital importance for space weather science and applications. In addition, fast CMEs are usually accompanied by long-duration eruptive solar flares and large solar energetic particle (SEP) events. Quantitative measurements of the physical properties of CMEs constitute an essential piece to completing the ``system science'' of solar eruptions, and hence are of fundamental importance for solar- and helio-physics. Last but not least, the understanding of solar CMEs serves as a ``stepping stone'' for addressing exo-space weather in other stellar systems and the habitability of exoplanets.

To make progress in achieving a comprehensive picture, the following measurements must be obtained:

\begin{itemize}
    \item The magnetic field configuration of the pre-eruption magnetic flux ropes and quantifying their plasma environment. This measurement is crucial in determining the formation and initiation conditions of CMEs.
    \item Measuring the size, mass, speed, and magnetic field of CMEs from their early eruption in the low corona to their propagation throughout the interplanetary space. These observational inputs are essential for predicting the arrival of CMEs and evaluating their geospace impacts.
    \item Mapping the morphology, magnetic field configuration, and plasma parameters of CME-driven shocks. This requirement is compulsory to study the origin of large SEP events, which are themselves important space weather manifestations. 
\end{itemize}

Traditionally white-light coronagraph observations have been the primary means for studying CMEs. However, radio observations provide a rich variety of complementary diagnostic methods for obtaining such observational constraints. We refer interested readers to recent reviews by \citet{Vourlidas2020} and \citet{Carley2020b} for more comprehensive discussions on the details of these methods and the progress made in the past two decades, as well as their complementarity to observations at other wavelengths. For a discussion of coronal cavity magnetic diagnostics at UV and optical/IR wavelengths, see the white paper by \citet{Gibson2022}. In this white paper, we will focus on radio diagnostics for coronal cavities and erupting CMEs, with an emphasis on the uniqueness in measuring the magnetic field and energetic electrons in CMEs and CME progenitors, as well as their highly complementary nature to optical/infrared spectropolarimetry techniques. We argue that the next major advance in radio diagnostics calls for a solar-dedicated instrument that is capable of performing broadband radio dynamic imaging spectroscopy and polarimetry with a high dynamic range, high image fidelity, and high sensitivity. 

\section{Coronal Cavities as CME Precursors}\label{sec:cav}

Coronal cavities have been observed in the Sun’s corona for several decades. They are of particular interest because of their close relationship with CMEs \citep{Gibson2006}.  CMEs observed in white light often show a three‐part structure: a bright expanding loop followed by a dark cavity containing a bright core that corresponds to an erupting prominence. This structure is mirrored in coronal cavities observed on the solar limb where a bright loop or coronal streamer and dark cavity are seen in quiescence; a bright prominence core is also commonly seen. Seen against the solar disk, coronal cavities overlie solar filaments or, more generally magnetic polarity inversion lines. A coronal cavity is often destabilized leading to a CME and hence they can be considered as CME precursors. These ejections are magnetically driven, the energy stored in twisted or sheared magnetic fields prior to eruption. %This being the case, \textbf{an understanding of the magnetic field in coronal cavities is critical to understanding CME initiation and energetics.}

%The central role of the magnetic flux ropes is further corroborated by \textit{in situ} measurements of interplanetary CMEs, which can be accounted for in many cases by an erupting flux rope model (\textbf{refs needed}). Such models have shown that a twisted flux rope morphology in a coronal cavity can provide the magnetic support needed by the prominence and could provide the free energy needed to drive the CMEs.
 
Clearly, \textbf{a quantitative understanding of the physical attributes of coronal cavities is needed, particularly of the magnetic field topology and the magnetic free energy available.} Significant progress has been made on the temperature and density structure of coronal cavities over the past decade. For example, \citet{Reeves2012} used the X-ray telescope (XRT)  on  board  the  Hinode  spacecraft  to study the thermal soft X-ray emission in cavities with hot cores, showing cavity temperatures of 1.5--1.65 MK and core temperatures of 1.7--2 MK. Forward-models of data from the Hinode EUV imaging Spectrometer (EIS) and the white-light polarization brightness measured by the Mauna Loa Solar Observatory (MLSO) Mk4 coronagraph were used \citep{Schmit2011} to infer the density of a cavity and the surrounding streamer from 1.05--1.25 $R_{\odot}$. In cavities, densities ranging from $0.5-3 \times  10^8$ cm$^{-3}$  were found over this radial range whereas streamer densities of $0.8-5 \times 10^8$ cm$^{-3}$ were found. For the  magnetic  field,  recent  results  from  the  Coronal  Multi-Channel Polarimeter (CoMP) show that the linearly polarized infrared (IR) emission in coronal cavities is consistent with a magnetic flux rope topology. However, \textbf{a critical missing ingredient is quantitative measurements of the vector magnetic field.}

At optical/IR wavelengths, spectropolarimetry observations have been used to constrain the plane-of-the-sky direction of the magnetic field of coronal cavities (based on the saturated Hanle effect; see, e.g., \citealt{2013ApJ...770L..28B}).  To measure the line-of-sight (LOS) component of the field, much more sensitive spectropolarimetry measurements are necessary to detect the extremely weak Stokes-V signal due to Zeeman splitting. The expected circular polarization degree (V/I) of the off-limb measurements is only $\lesssim0.1$\%, or $10^{-9}$ of the brightness at the disk center. As a result, such measurements are only possible for off-limb sources and require large-aperture coronagraphs operating at a telescope site with excellent atmospheric conditions. The Daniel K. Inouye Solar Telescope (DKIST) and the proposed COronal Solar Magnetism Observatory (COSMO) are poised to make major advances in this area (see the white paper by \citealt{Tomczyk2022} for details). 

At radio wavelengths, the Stokes-V signal of the coronal magnetic field via thermal free-free emission is, in fact, prominent. In the optically thin regime, the degree of circular polarization $p=V/I$ is given by
$p \approx [0.56 B_{\rm LOS} / \nu_{\rm GHz}]$\%, where $B_{\rm LOS}$ is the LOS component of the magnetic field (weighted by the square of the density in the source) and $\nu_{\rm GHz}$ is the observing frequency in GHz \citep{Casini2017}.
For example, a $B_{\rm LOS}$ of 20 G would give rise to a polarization degree of order 10\% at 1 GHz---that is, 100 times stronger than that of the optical/IR signal via Zeeman splitting! Thanks to their comparable radio brightness relative to the disk, \textbf{such $B_{\rm LOS}$ measurements can be made for both the coronal cavities above the limb and, uniquely, also their on-disk counterparts---sigmoids or filament channels}. In addition, 
%because of the relatively high brightness temperature ($T_b$) of the radio source---10$^4$--10$^6$ K, compared to a system-noise-equivalent ``temperature'' of a few $\times$100--1000 K for a typical solar radio telescope---s
such measurements can be obtained at a time cadence of few seconds. Therefore, \textbf{the radio measurements can be used to reveal rapid dynamic evolution of the plasma structure and magnetic field configuration of CME precursors}, which is another unique diagnostic potential available to this technique.

\begin{figure}[!ht]
%\floatbox[{\capbeside\thisfloatsetup{capbesideposition={right,top},capbesidewidth=4.3cm}}]{figure}[\FBwidth]
\includegraphics[width=1.0\textwidth]{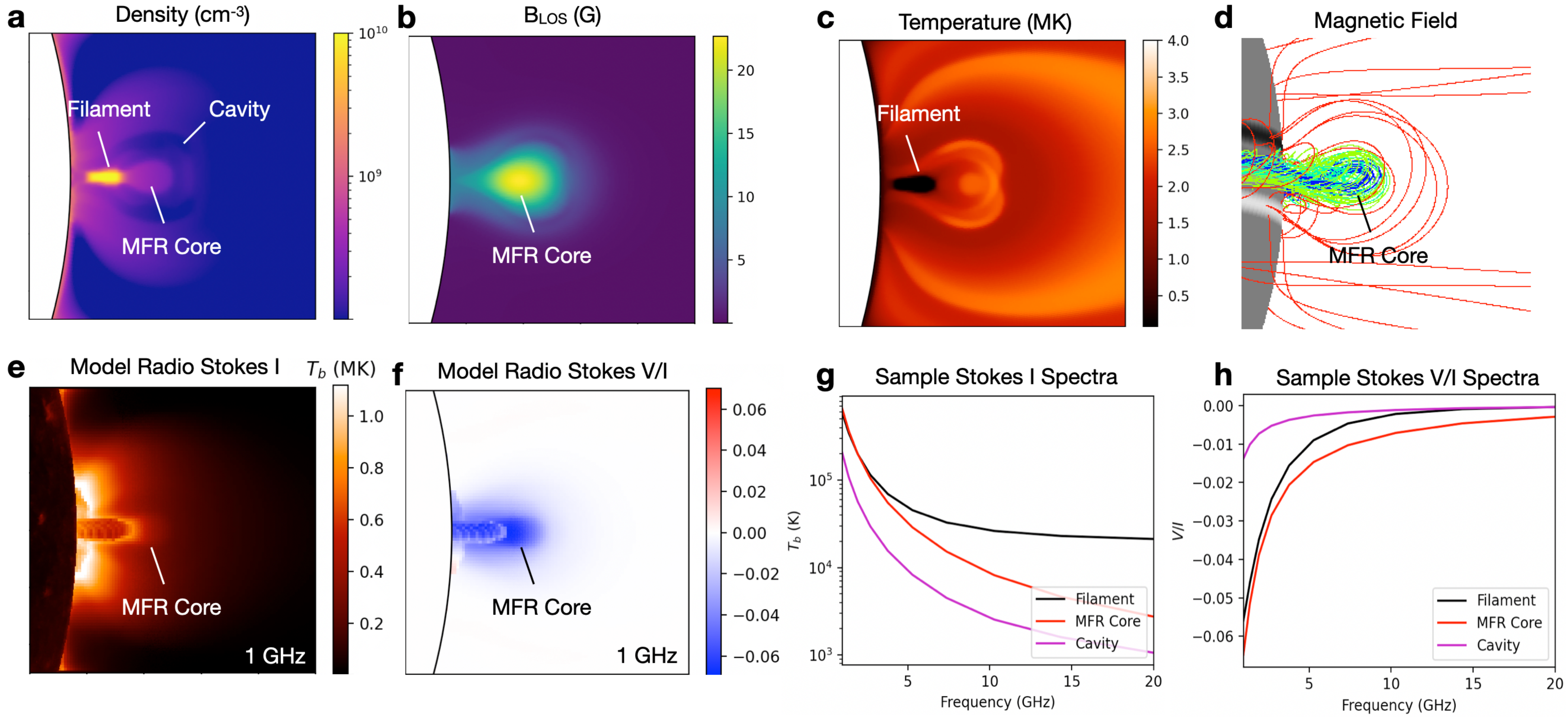}
\caption{\fontsize{11}{13}\selectfont Radio observations can be used to derive maps of key physical parameters of coronal cavities, including their plasma density, temperature, and LOS magnetic field. (a--d): Parameters of the 3D MHD coronal cavity model (showing those at the mid-plane), which features a filament-hosting twisted magnetic flux rope viewed end on (from \citealt{Fan2019}). (e) and (f): Synthetic radio thermal free-free Stokes I and V/I maps at 1 GHz calculated by applying radiative transfer to the 3D MHD model. They bear striking similarities to the density and $B_{\rm LOS}$ distributions in the MHD model. (g) and (h): Stokes I and V/I spectra sampled at three locations that correspond to the filament, flux rope core, and cavity. Note the core of the flux rope features a strong Stokes V signal of $\sim$1--7\% over a wide range of frequencies from 1--20 GHz.}\label{fig:cavity_model}
\end{figure}

As a demonstration, we use a realistic 3D MHD model of a magnetic-flux-rope-hosting coronal cavity developed by \citet{Fan2019} (Fig. \ref{fig:cavity_model}) to calculate the corresponding synthetic radio Stokes I and V emission maps at multiples frequencies using the \texttt{FORWARD} package \citep{Gibson2016}. An example for 1 GHz is shown in Figs. \ref{fig:cavity_model}(e) and (f). For this model, the Stokes V/I signal at 1 GHz reaches a peak of $\sim$7\% in a region that coincides with the core of the flux rope (located just above the dense and cool filament) where $B_{\rm LOS}$ reaches a maximum. Multi-frequency mapping enables spectral analysis (panels g, h), and the spatial distribution and temporal variation of the Stokes V and I maps serve as an excellent tool to derive the spatial and temporal evolution of key physical parameters including $B_{\rm LOS}$, plasma density, and temperature.

Despite the potential of radio observations for deriving the detailed morphology of coronal cavities (both on the limb and against the disk) and accurately mapping $B_{\rm LOS}$, \textbf{current radio facilities do not have the adequate imaging fidelity, dynamic range, and polarization purity necessary to make such measurements.} The figure on the cover page shows one of the best examples of multi-frequency radio images of a coronal cavity obtained by the Expanded Owens Valley Solar Array (EOVSA), a key pathfinder instrument for demonstrating these techniques. At low frequencies (e.g., 1.4 GHz), the filament-cavity system features a dark cavity with a bright rim. This is because the optical depth in the low-density cavity is small compared to its rim. At higher frequencies (e.g., 13.9 GHz), however, the optically thick emission from the cooler but much denser filament starts to dominate the emission. These multi-frequency images demonstrate the feasibility of imaging the filament-cavity systems at radio wavelengths and, more importantly, the potential of using spectral analysis to derive the plasma parameters. However, owing to the small number of baselines from EOVSA's 13 antennas, 
these images can only be made with several hours of integration by taking advantage of Earth's rotation to improve the u-v coverage. Thus, crucial details of the \textit{dynamic evolution} of the coronal cavity structures are not recovered. Additionally, owing to its limited spatial resolution, image fidelity, and dynamic range, measuring the Stokes-V signal down to the required detection level of a few percent is likely out of reach for EOVSA.

\begin{figure}[!ht]
%\floatbox[{\capbeside\thisfloatsetup{capbesideposition={right,top},capbesidewidth=4.3cm}}]{figure}[\FBwidth]
\includegraphics[width=0.9\textwidth]{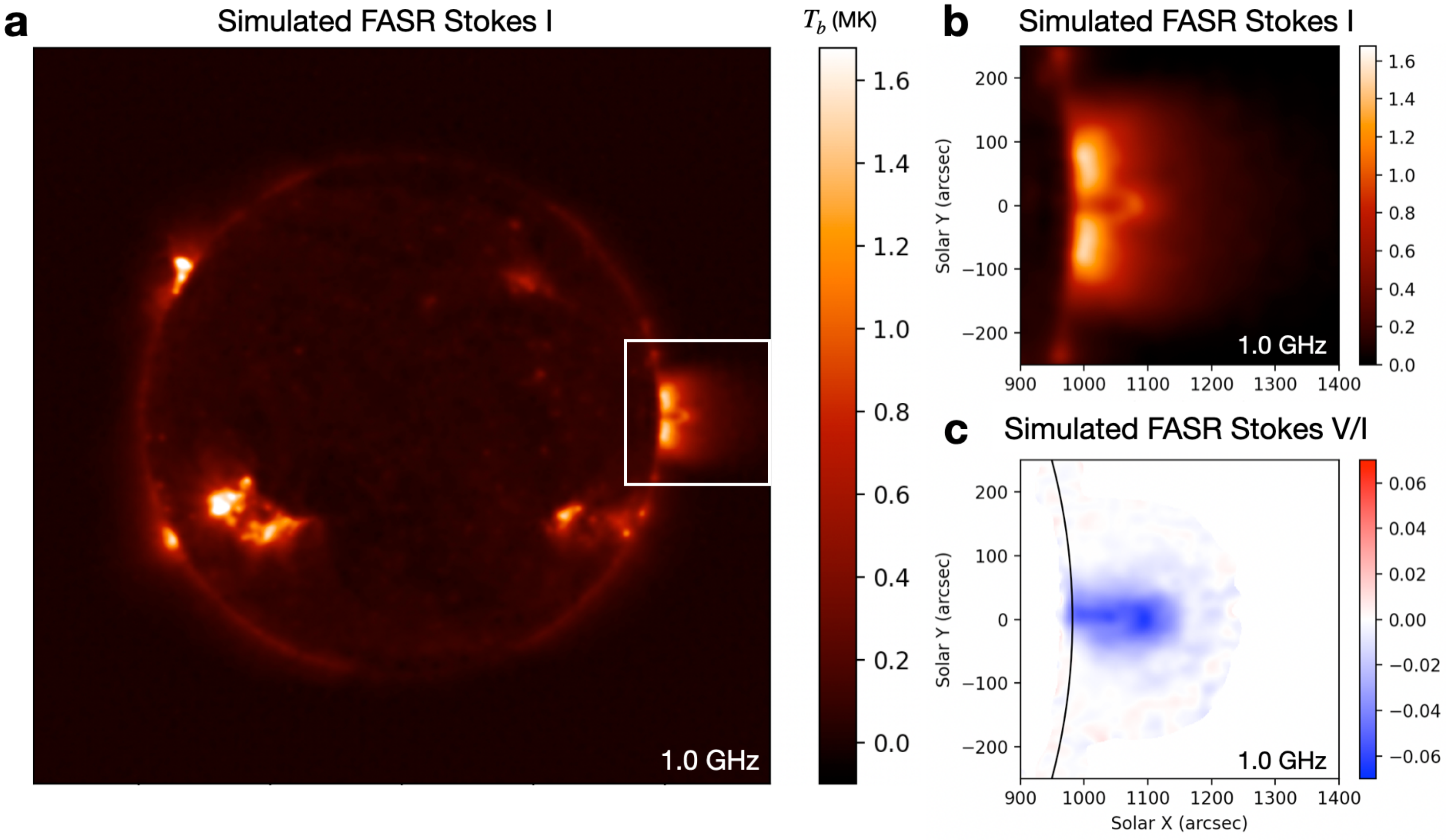}
\caption{\fontsize{11}{13}\selectfont Simulated snapshot (1-s cadence) Stokes I and V images by an example FASR array configuration that consists of 200 2-m-diameter antennas. (a) Simulated full-disk image at 1 GHz. (b) and (c) Enlarged view of the Stokes I and V/I images of the coronal cavity region (white box in (a)). The latter is simulated from an input emission model calculated from the 3D MHD model shown in Fig. \ref{fig:cavity_model}. The angular resolution of the array is $\sim$20$''$/GHz. See text for details.}\label{fig:cavity_fasr}
\end{figure}

To fully exploit the potential of these radio diagnostics to study CME precursors, a radio array with a large number of antenna elements (of order 100) and accurate spectropolarimetry capabilities over a frequency range of $\sim$0.5--20 GHz is required. With the $>$2 orders of magnitude increase in the number of baselines sampling the u-v domain compared to, e.g., EOVSA, the instrument will provide dramatically improved image fidelity and dynamic range. Such a superior imaging capability would allow all features with different angular scales and brightness across the full-Sun disk to be faithfully recovered. Fig. \ref{fig:cavity_fasr} shows simulated snapshot (1-s cadence) imaging results based on an example 200-element log-spiral array configuration (see Fig. 3 in \citealt{Chen2022a} for plots of the array configuration and performance metrics). In order to best emulate the actual observational conditions, the input ``sky model'' is made from the synthetic radio emission maps calculated from the 3D MHD coronal cavity model, as shown in Fig. \ref{fig:cavity_model}, combined with a synthetic full-disk radio map. The latter is constructed by calculating the thermal free-free radio emission using the differential emission measure map derived from multi-band SDO/AIA EUV images \citep{Fleishman2021}. This sky model is then ``observed'' with the model array configuration to produce complex visibilities using CASA's \texttt{simobserve} task, and is then subsequently Fourier transformed and deconvolved using the multi-scale \texttt{CLEAN} algorithm \citep{Cornwell2008}. Random noise (due to the instrument's system temperature and residual gain errors after calibration) is introduced in the simulation to best replicate the actual observations. 

Contrasting the ``ground truth'' shown in Fig. \ref{fig:cavity_model} with the simulated observations, it is evident that crucial details of the cavity brightness, morphology, and Stokes-V signatures are faithfully reproduced. Therefore, with the added broadband multi-frequency capability for performing spatially resolved spectral analysis (based on spectra similar to those in Fig. \ref{fig:cavity_model}(g) and (h)), \textbf{such a next-generation instrument is expected to make a giant leap forward in constructing the key parameters of coronal cavities, including the elusive measurements of the strength, spatial distribution, and short-time-scale evolution of the LOS coronal magnetic field.} We emphasize that, although the discussion above focuses on coronal cavities that are long-lived structures typically observed in quiet Sun regions, with the second-scale snapshot imaging spectropolarimetry capability, \textbf{similar techniques can be applied to fast-evolving, flux-rope-hosting structures in active regions.} These active region flux rope structures, which can lead to major eruptive flare events and fast CMEs, are expected to yield more significant radio signatures because they are hotter, denser, and bears a stronger magnetic field.

%\section{Radio Observations of Flux Rope Eruption}\label{sec:fr}

%\citep{Chen2014, Chen2015, Chen2017, 2018ApJ...863...83G}:nonthermal signatures of the flux rope eruption: thermal emission with sufficient DR? Previous AIA work. Nonthermal GS and coherent radio diagnostics of underlying loops, reconnecting current sheet, termination shock, etc. 

\vspace{-0.3cm}

\section{Radio Observations of CMEs and CME-driven Shocks}\label{sec:cme}

When CMEs erupt, in addition to the radio emission from the ejected filaments and CME bubbles \citep[e.g.,][]{Huang2019}, energetic electrons accelerated or trapped by the CMEs or CME-driven shocks also emit a variety of nonthermal radio emissions. 

\subsection{Type II Radio Bursts}

The most common radio signature of a fast CME is a type II radio burst, coherent plasma radiation produced by the CME-driven super-Alfv\'{e}nic shock propagating in the corona and sometimes well out into the interplanetary medium. Type IIs are recognized by their characteristic slow-drift signature in dynamic spectra and have been used to constrain the density of the medium and the shock speed. They have been imaged at discrete frequencies in past years \citep[e.g.,][]{Bain2012} but their relation to the underlying CME driver has been murky. In more recent years, instruments like the general-purpose LOFAR and the MWA have produced multi-frequency imaging of type IIs 
%and the associated fine structures such as herringbones (interpreted as escaping electron beams; 
\citep{Morosan2019}. Determining the relationship between type II radio bursts and the shock driver is a key science goal of solar-dedicated imaging arrays being commissioned at long wavelengths, which include Owens Valley Long Wavelength Array \citep[OVRO-LWA, 20-88 MHz;][]{chhabra2021} and the SunRISE mission \citep{Kasper2022}, which will use a space-based interferometric array covering 0.1--25 MHz. However, \textbf{tracking the initiation and development of type II radio bursts requires continuous coverage from the low corona to the middle corona.} Under typical coronal conditions, this translates to a frequency range from $\sim\!500$ MHz down to $<$20 MHz. As reviewed by \citet{Chen2022b}, currently \textbf{there is no solar-dedicated radio instrument available that provides broadband dynamic imaging spectroscopy over this frequency range, although LOFAR, the MWA, and OVRO-LWA partially address the need.} 
%which corresponds to a key region where the primary CME acceleration and shock development occurs} \citep{West2022}. Therefore, a solar-dedicated instrument with such capabilities is critically needed.

\subsection{Radio CMEs}

Faint radio emissions that closely resemble their white light CME counterparts are dubbed ``radio CMEs'' because of their similar appearance \citep[see a recent review by][]{Vourlidas2020}. %In fact, they were discovered around the same period as LASCO's start of science operations in 1996. 
The emission mechanism is believed to be synchrotron emission which,
%, which occurs at large harmonics of the electron gyrofrequency and is thus well above the local plasma frequency, thereby being less affected by scattering. Similar to the free-free emission from coronal cavities discussed above, thanks to their incoherent nature, 
when imaged spectroscopically, can be used to map the evolving CME magnetic field strength (and possibly direction, if polarimetry is available), non-thermal electron distribution, and the thermal electron number density \citep[see, e.g.][]{Bastian2001}. It should be noted that the same plasma producing white light CMEs in the middle to high corona also emits \textit{thermal} free-free radiation. This is usually difficult to detect due to low surface brightness (see discussion by \citealt{Bastian1997}), although detections have been reported in the literature \citep{Gopalswamy1993,Ramesh2021}.

\begin{figure}[!ht]
\floatbox[{\capbeside\thisfloatsetup{capbesideposition={right,center},capbesidewidth=4.2cm}}]{figure}[\FBwidth]
{\includegraphics[width=0.6\textwidth]{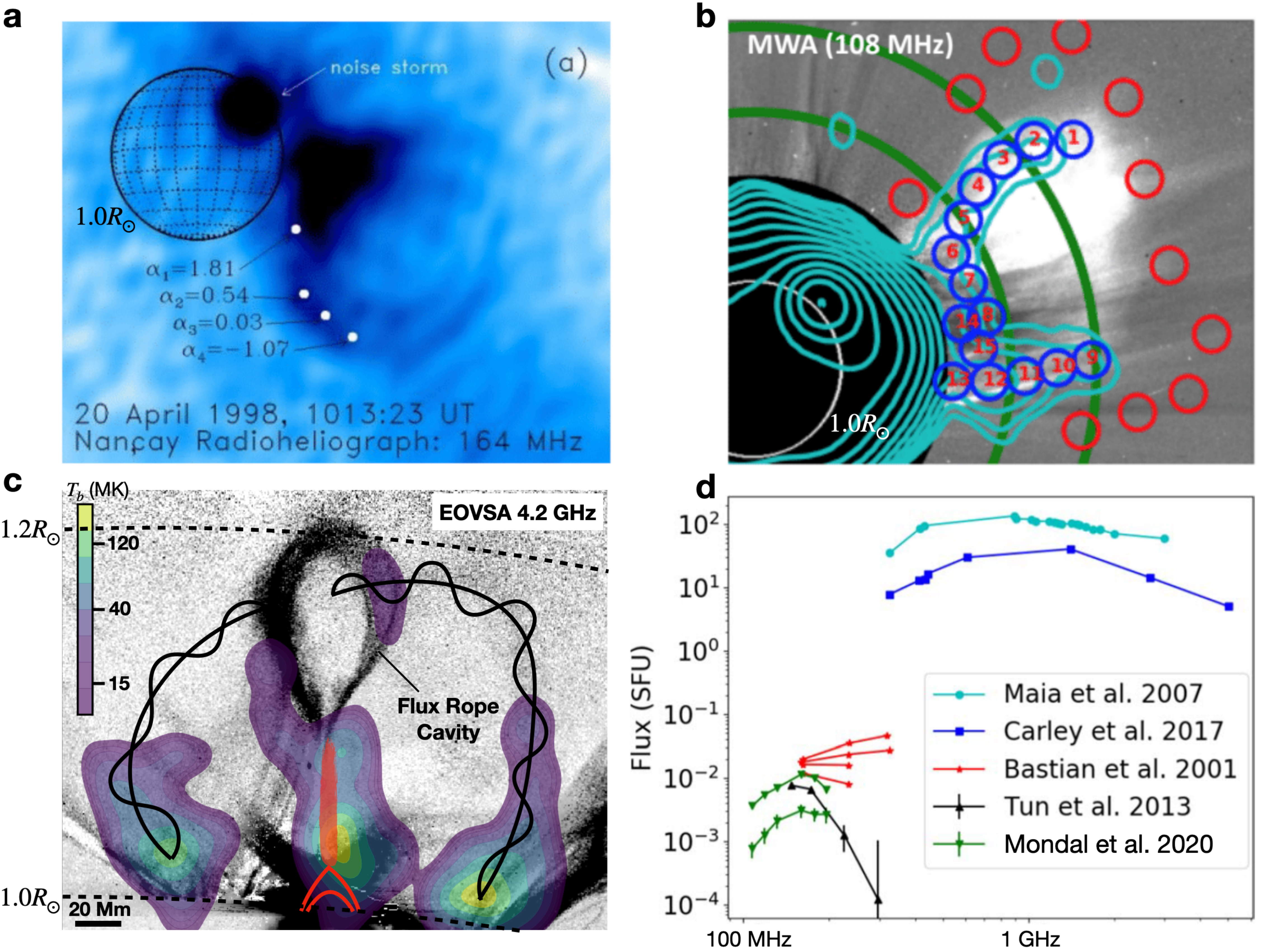}}
{\caption{\fontsize{11}{13}\selectfont Examples of ``radio CME'' observations by the Nan\c{c}ay Radioheliograph \citep{Bastian2001} (a) and the MWA \citep{mondal2020} (b) in the middle corona at metric wavelengths. (c) A high-frequency counterpart of radio CME observed by EOVSA in the low corona at $\sim$4.2 GHz \citep{Chen2020b}. (d) Spectra of some reported radio CME events \citep{mondal2020}.}\label{fig:cme}}
\end{figure}

Since the original detections made by the Nan\c{c}ay Radioheliograph operating between 150--450 MHz (see Fig. \ref{fig:cme}(a) for an example), few radio CMEs have been reported. Recently, emission surrounding a flux rope cavity, which could be considered a radio CME at its initiation phase, was detected with EOVSA at $\sim$4.2 GHz in the low corona ($\sim\!1.2R_{\odot}$; Fig. \ref{fig:cme}(c)). %To our knowledge, almost all the detected radio CMEs are fast CMEs with a speed $>$1,000 km/s. 
More recently, 
%using the Murchison Widefield Array (MWA), a 128-antenna array that provides highly sensitive, high dynamic range imaging, 
a faint radio CME event was observed from 80--240 MHz by the MWA (Fig. \ref{fig:cme}(b); \citealt{mondal2020}). The event shared a similar morphology to those reported earlier but was associated with a much slower ($<$500 km s$^{-1}$) and fainter (of order 0.001--0.01 sfu) CME. Such an unprecedented detection was made possible thanks to MWA's ability to perform highly sensitive, high dynamic range imaging with its 128 antenna elements. With spatially resolved spectral analysis, a magnetic field strength of $\sim$10 G was derived at the outer ``shell'' of the CME bubble. In addition, the number density and power-law index of the nonthermal electron distribution were also constrained.% to have a total number density of $\sim\,10^4\,\mathrm{cm}^{-3}$ above 3 keV and a power-law index of 3--4. 

This outstanding example, obtained by a state-of-the-art general-purpose facility (i.e. proposal driven), provides a glimpse of the unique power of radio observations for constraining the magnetic field configuration and nonthermal electron distribution of erupting CMEs with a variety of speeds and strengths. \textbf{These diagnostics, however, call for routine solar-dedicated observations with high sensitivity and high dynamic range imaging spectroscopy. 
In addition, to trace and measure the radio CMEs from their initiation in the low corona to their evolution in the middle corona, a continuous frequency coverage over 10s of MHz to a few GHz is required.} 
%over a frequency range similar to that required to observe CME-driven type II radio bursts: several 10s of MHz to several times 100 MHz.}

%A paragraph on FR observations of ICMEs using JVLA and later, ngVLA. 

\section{Observing the System: Findings and Recommendations}

\vspace{-0.1cm}

To conclude, radio observations can be used in a unique and wholly new manner to constrain the formation, evolution, destabilization, and eruption of CMEs and CME progenitors. Especially powerful will be joint optical/IR/(E)UV and radio observations of coronal cavities as \textit{CME precursors} and their destabilization into nascent CMEs, with multi-wavelength coronal spectropolarimetry yielding sensitivity to different parts of the emitting coronal plasma through a variety of physical mechanisms  (\citealt{Gibson2021}; see e.g., Table 1), leading to the robust reconstruction of their 3D coronal magnetic field. Equally important are the \textit{unique} diagnostics offered by radio observations on the magnetic field and nonthermal electron distribution of \textit{erupting CMEs and the associated shocks} as they lift off and propagate into the upper corona. To briefly summarize the discussion above:

\vspace{-0.1cm}

\begin{itemize}
\setlength\itemsep{0em}
\item{CME precursors (Sec. \ref{sec:cav}): Radio (this paper) and O/IR (white paper by \citealt{Gibson2022}) observations from the ground and UV observations from space (white paper by \citealt{Casini2022}) offer powerful and complementary diagnostics of the plasma environment in and around coronal cavities. 
%on the solar limb, constraining the plasma environment and, in particular, the LOS magnetic field. 
Radio observations of thermal free-free emission and polarimetry will yield the line-of-sight component of the magnetic field of coronal cavities and flux-rope-hosting structures in both quiescent and active regions. Unlike optical/IR observations, which for this goal can only be taken above the limb, radio observations can be made of coronal cavities, filament channels, and sigmoids on the disk.}
%\item Flux rope eruption: Radio and O/IR observations from the ground will also be a strong combination for observing the destabilization of flux ropes and their acceleration up into the corona. Radio observations are exquisitely sensitive to nonthermal particles and will indicate when and where they occur as the flux rope erupts (Sec. \ref{sec:cme}). As the eruption evolves into an eruptive flare, the magnetic field and the electron distribution function can be measured as a function of time and space -- in the reconnecting magnetic loops, the current sheet, and the flux rope. %Coherent emissions may yield insights into magnetic energy release and a termination shock.
\item{Erupting CMEs and CME-driven shocks (Sec. \ref{sec:cme}): The erupting magnetic flux ropes/CMEs can be imagined at radio wavelengths if it has significant numbers of energetic electrons entrained in their magnetic field. As the eruption evolves into an eruptive flare, the magnetic field and the electron distribution function can be measured as a function of time and space---in the reconnecting magnetic loops/current sheet (white paper by \citealt{Chen2022a}) and the erupting CME themselves (this paper). Ground-based radio observations can make these measurements dynamically across the evolving CME out to several solar radii. Faraday Rotation (FR) measurements provide additional constraints on the CME density and magnetic field \citep{Kooi2022wp}. Any type II radio burst, tracing a shock that accompanies the CME, may also be imaged simultaneously.}
\end{itemize}

%In addition, as the CME propagates into the IPM, when passing in front of polarized background sources (see the white paper by \citealt{Kooi2022}) or between space-based transmitters and receivers (see white paper on the MOST mission by N. Gopalswamy et al.), Faraday Rotation (FR) measurements provide additional constraints on the CME density and magnetic field. Ground-based FR observations require extremely sensitive instruments like the Jansky Very Large Array \citep[e.g.,][]{Kooi2021} or the proposed ngVLA while space-based FR observations will require advanced instruments like the FETCH instrument on board the MOST science mission.

%Radio techniques then provide entirely unique observables of the eruptive flares and CMEs. 

\textbf{The above elements require radio coverage from a few 10s of MHz to 20 GHz in order to observe them as a coupled system.} To make transformative advances in making new and unique measurements discussed above, a solar-dedicated radio facility with a large number of elements (of order 100) is required to obtain broadband radio imaging spectropolarimetry with high image fidelity, dynamic range, angular resolution, and sensitivity. In the past decade, significant investments have been made in ground-based low-frequency arrays at frequencies $\lesssim$0.2 GHz with the operation of LOFAR, MWA, and the solar-dedicated OVRO-LWA (see Fig. 5 of \citealt{Chen2022b} for a synopsis). Therefore, in the coming decade, we suggest the focus should be placed in the $\sim$0.2--20 GHz range.

The \textit{Frequency Agile Solar Radiotelescope} (FASR) is a concept for realizing such high sensitivity, image fidelity, and dynamic range imaging spectropolarimetry observations in an ultra-wide frequency range of $\sim$0.2--20 GHz. We refer readers to \citet{Gary2022c} and references therein for a more detailed discussion of the instrument and a broad range of breakthrough science it is expected to achieve.

\bibliography{references}
\bibliographystyle{aasjournal_bc}

\end{document}